Publication date: July 2023
Embargo: 24 Months
European Union, Horizon 2020, Grant Agreement number: 857470 — NOMATEN — H2020-WIDESPREAD-2018-2020
DOI: https://doi.org/10.1016/j.nimb.2023.03.027

# High versus low energy ion irradiation impact on functional properties of PLD-grown alumina coatings

A. Zaborowska[a], Ł. Kurpaska[a,*], E.Wyszkowska[a], A. Azarov[b], M.Turek[c], A.Kosińska[a], M. Frelek-Kozak[a], J. Jagielski[a,d]

[a] National Center for Nuclear Research, A. Soltana 7, 05-400 Otwock-Swierk, Poland
[b] Centre for Material Science and Nanotechnology, University of Oslo, P.O. Box 1048, Blindern, N-0316 Oslo, Norway
[c] Institute of Physics, Maria Curie Skłodowska University, pl. M. Curie-Skłodowskiej 1, 20-031 Lublin, Poland
[d] Lukasiewicz Institute for Microelectronics and Photonics, Al. Lotnikow 32/46, 02-668 Warsaw, Poland

*corresponding author: lukasz.kurpaska@ncbj.gov.pl, Phone: (+48) 22 273 1061, Fax: (+48) 22 273 1061

Abstract

It is well known that ion irradiation can be successfully used to reproduce microstructural features triggered by neutron irradiation. Even though the irradiation process brings many benefits, it is also associated with several drawbacks. For example, the penetration depth of the ion in the material is very limited. This is particularly important for energies below MeV, ultimately reducing the number of available irradiation facilities. In addition to that, extracting information exclusively from the modified volume may be challenging. Therefore, extreme caution must be taken when interpreting obtained data. Our work aims to compare the findings of nanomechanical studies already conducted separately on thin amorphous ceramic coatings irradiated with ions of different energies, hence layers of different thicknesses. In this work, we show that in some instances, the 10% rule may be obeyed. In order to prove our finding, we compared results obtained for ion irradiated (with two energies: 0.25 and 1.2 MeV up to 25dpa) alumina coating system. Mechanical properties of pristine and ion-irradiated specimens were studied by nanoindentation technique. Interestingly, the qualitative relationship between nanohardness and irradiation damage level is very similar, regardless of the energy used. The presented work proves that for some materials (e.g., hard coatings), the qualitative assessment of the mechanical changes using nanoindentation might be feasible even for shallow implantation depths.

**Accepted Version**

Publication date: July 2023
Embargo: 24 Months
European Union, Horizon 2020, Grant Agreement number: 857470 — NOMATEN — H2020-WIDESPREAD-2018-2020
DOI: https://doi.org/10.1016/j.nimb.2023.03.027

1. Introduction

One potential application of alumina is utilization as a coating for fuel cladding material in Generation IV nuclear concepts, e.g. lead-cooled fast reactor (LFR) [1]. Conditions imposed by this advanced concept on in-core structural materials include high temperature (550 °C [2]), very high level of radiation damage (up to 100 dpa [2]), and a very harsh working environment (heavy liquid metal [2]). As of today, $Al_2O_3$ coatings have been manufactured by several methods, including chemical vapor deposition, physical vapor deposition, spray techniques, sol-gel process, and electrochemical processing [3–5]. The properties of the coating strongly depend on the deposition technique and temperature [2]. Pulsed laser deposition (PLD) is one of the physical vapor deposition (PVD) techniques that can be used to grow thin alumina films [4,6–8] and permits the deposition at low substrate temperatures [9]. In the past decade, a growing interest of the nuclear community in PLD-grown alumina coatings has been observed [1,3,9–17].

In the work of F. García Ferré et al. [14], the group investigated the mechanical properties of the coatings grown by PLD at either RT or 600 °C. Conducted studies revealed that the coating grown at RT is composed of homogeneously dispersed ultrafine γ-$Al_2O_3$ nanodomains in an amorphous matrix and exhibit excellent mechanical performance like high hardness (H = 10.0 ± 0.3 GPa), moderate stiffness (E= 195 ± 9 GPa and υ = 0.29 ± 0.02), superior plastic behavior (over other ceramic coatings), remarkable wear resistance and strong interfacial bonding. The combination of all these features coupled with chemical inertness and high-temperature resistance is specific to PLD-grown alumina-based coatings [13,15]. For this reason, these materials are considered promising for nuclear applications where radiation and corrosion resistance at high temperatures are mandatory conditions. F. García Ferré and co-workers, in their report [12], investigated the evolution of the microstructure, mechanical properties, and impact resistance of $Al_2O_3$ thin films subjected to ion irradiation at 600 °C. For the first time, published data clearly showed that irradiation induces crystallization of the initially

**Accepted Version**

Publication date: July 2023
Embargo: 24 Months
European Union, Horizon 2020, Grant Agreement number: 857470 — NOMATEN — H2020-WIDESPREAD-2018-2020
DOI: https://doi.org/10.1016/j.nimb.2023.03.027

amorphous alumina phase. As crystalline alumina yields higher hardness (~25 GPa) than amorphous (~10-12 GPa) [18], the hardness of the material increases as a result of phase transformation. Further increase of the damage level is accompanied by grain growth and softening effect, which follows the Hall-Petch relationship [12]. The most important conclusion from this study is that PLD-grown alumina-based coating exhibits excellent radiation resistance at high temperatures. New important insights into alumina irradiation resistance phenomena and corrosion behavior were also provided by F. García Ferré et al. [15]. The research results showed that PLD alumina effectively prevents Pb corrosion of the steel substrate, regardless of the sample condition (i.e., virgin or irradiated).

Briefly presented properties of the alumina coatings prove their importance to new industries where extreme operational conditions are to be used. One of these potential industries is the nuclear business. As demonstrated above based on the alumina example, materials devoted to nuclear applications must be properly qualified, and their behavior in operating conditions has to be fully understood. Ion irradiation is broadly used for the purpose - to emulate neutron radiation damage [19–22]. The benefits gained from the technique are high damage rates, strict control of irradiation conditions, little or no material activation, and the resulting low costs of post-irradiation characterization (compared to the utilization of a Hot Cell facility) [19,23]. Another one important reason why ion irradiation has a significant advantage over neutron irradiation is that the availability of ion beam accelerators compared to test reactors is significantly higher. On the other hand, it is still insufficient to cover the needs arising in the nuclear community. Easy, flexible and efficient access to ion irradiation facilities with high capabilities is very limited for many researchers. In fact, many of them have only access to ion beam facilities that can only provide low-energy ions.

At the same time when using nanoindentation on low-energy (thus shallowly) irradiated material, the issue of indentation depths for which extracted mechanical property information reflects properties of ion irradiated layer becomes significant [24]. The upper end of the range is due to the high contribution of signal from the non-irradiated substrate, and the lower is due to the not fully developed plastic zone in the material. In

**Accepted Version**
Publication date: July 2023
Embargo: 24 Months
European Union, Horizon 2020, Grant Agreement number: 857470 — NOMATEN — H2020-WIDESPREAD-2018-2020
DOI: https://doi.org/10.1016/j.nimb.2023.03.027

certain cases, these conditions are mutually exclusive, and mechanical property information gained is affected by the effects described. The aim of this study is to verify whether such data may still be valuable when evaluating the influence of ion irradiation on material properties.

For this purpose, we analyzed the data collected from works [16,25] where thin films deposited on 316L SS via the PLD technique were ion irradiated at room temperature with 250 keV and 1.2 MeV $Au^+$ ions up to the dose of 25 dpa. Afterward, post-irradiation mechanical characterization of the samples was performed via the nanoindentation technique. The low-energy experiment presented here is a perfect example of a scenario described above when mechanical information is affected by a signal from non-modified volume and tip-rounding effect. At the same time high-energy experiment can be considered as free of these undesirable side effects. Nanomechanical results obtained for both experimental energies were compared. The comparison aims to verify whether the quantitative assessment of the mechanical changes occurring in the ion irradiated PLD-grown alumina using nanoindentation may be feasible even for very shallow implantation depths. Presented comparison sheds new light on the issue of very shallow ion irradiated material – especially regarding the detection limit. The findings of this work may be extended to other shallowly indented materials.

2. Material and methods

2.1. Materials preparation

1 μm $Al_2O_3$ coatings were deposited by Pulsed Laser Deposition (PLD) at room temperature on 1x1 cm stainless steel substrates. Substrates were cut from a 0.9 mm thick sheet of annealed AISI 316L steel purchased from Goodfellow. Details of the deposition process have been given elsewhere [14]. According to F. García Ferré et al., PLD-grown coatings generally reproduce the roughness of the substrate. Therefore roughness of the coating oscillated around 10-20 nm, which is similar to the roughness of the steel. Precise roughness measurement was done because it is known that a high roughness surface can considerably affect the nanoindentation results. Prior to the manufacturing process, steel

**Accepted Version**

Publication date: July 2023
Embargo: 24 Months
European Union, Horizon 2020, Grant Agreement number: 857470 — NOMATEN — H2020-WIDESPREAD-2018-2020
DOI: https://doi.org/10.1016/j.nimb.2023.03.027

samples were ground with SiC abrasive papers (up to 1200 grade) and finally mirror-polished with colloidal silica suspension (0.06 μm grain size).

2.2. Ion irradiation and SRIM calculations

As-deposited samples were irradiated at room temperature using two different energies – low (250 keV) and high (1.2 MeV). Irradiations were performed with $Au^+$ ions up to six fluences in the range of approx. $8.0 \times 10^{13}$ cm$^{-2}$ to $4.0 \times 10^{15}$ cm$^{-2}$. A separate sample was used for each combination of energy and fluence. Fluences were selected basing on the SRIM (The Stopping and Range of Ions in Matter) [26] simulations, and they correspond to damage levels of 0.5; 1; 3; 5; 10, and 25 dpa. The SRIM simulated damage and ion distribution profiles for irradiations carried up to highest fluence are shown in Fig. 1. Both damage profiles plotted with the same scale are shown in Fig. 2. One can clearly see that for low and high energy irradiations, produced radiation damage ranges are 70 and 270 nm, respectively. Selected irradiations conditions provide low ENSP (Electronic to nuclear stopping power) ratios, as suggested in [27]. Predictably, the ENSP profile for 250 keV irradiation is lower than for 1.2 MeV irradiation (at 20 nm – 1 vs 2.5). Low energy irradiations were carried out using the ion implantator UNIMAS-79 at the Institute of Physics, Maria Curie-Skłodowska in Lublin. High energy implantation campaign was performed at the Norwegian Micro- and Nano-Fabrication Facility, NorFab, on a 3SDH-2 NEC Pelletron Tandem.

**Accepted Version**
Publication date: July 2023
Embargo: 24 Months
European Union, Horizon 2020, Grant Agreement number: 857470 — NOMATEN — H2020-WIDESPREAD-2018-2020
DOI: https://doi.org/10.1016/j.nimb.2023.03.027

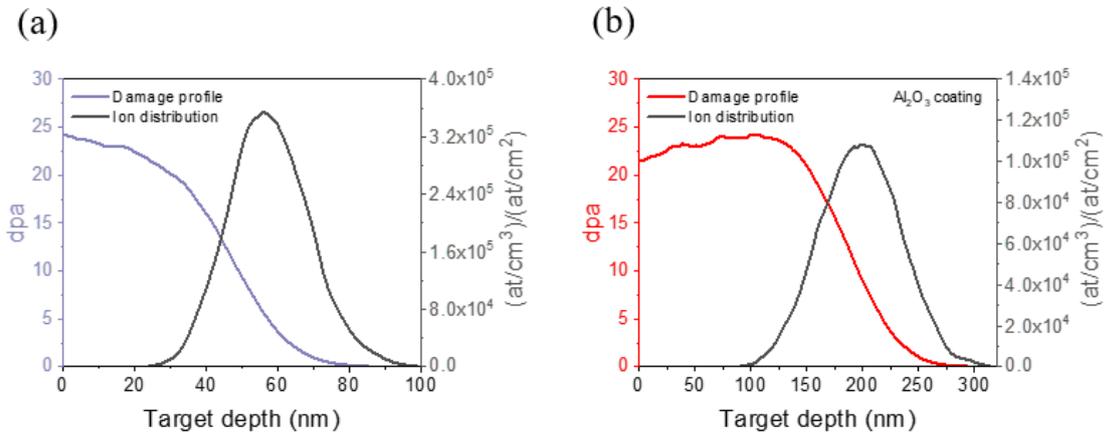

Fig. 1. Damage and ion distribution profiles for (a) 250 keV (reprinted from [16] with permission from Elsevier) and (b) 1.2 MeV Au$^+$ (reprinted from [25] with permission from Elsevier) irradiations of PLD-grown Al$_2$O$_3$ coatings up to 4.0 x 10$^{15}$ cm$^{-2}$

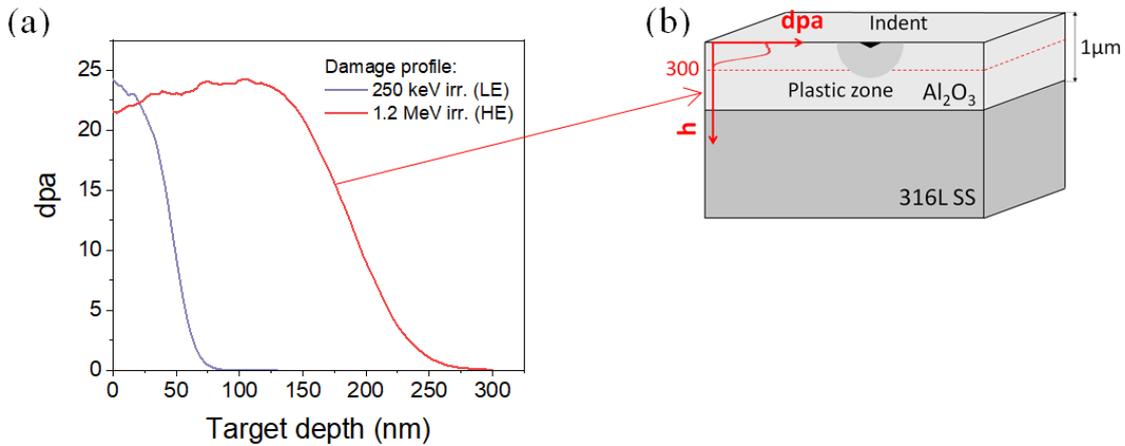

Fig. 2. (a) Comparison of damage profiles for 250 keV (blue line) and 1.2 MeV Au$^+$ (red line) irradiations in PLD-grown Al$_2$O$_3$ coatings up to 4.0 x 10$^{15}$ cm$^{-2}$; (b) Schematic representation of nanoindentation measurement on ion irradiated coating

2.3. SEM and roughness measurements

In order to examine the quality of the sample's surface, Scanning Electron Microscopy (SEM) observations and roughness measurements were performed. Three samples were selected for this investigation: (i) virgin, (ii) after low, and (iii) high energy

**Accepted Version**

Publication date: July 2023
Embargo: 24 Months
European Union, Horizon 2020, Grant Agreement number: 857470 — NOMATEN — H2020-WIDESPREAD-2018-2020
DOI: https://doi.org/10.1016/j.nimb.2023.03.027

irradiation up to 25 dpa. SEM observations were carried out using Hitachi SU8230 and Carl Zeiss Auriga CrossBeam Workstation. The device used to measure roughness of the specimens was the contact-type instrument Hommelwerke model LV-50. For each specimen, five measurements on 0.5 mm distance were performed, and the average roughness value $R_a$ was calculated.

2.4. Nanoindentation

The instrument used to perform nanomechanical measurements was NanoTest Vantage (Micro Materials, UK), equipped with a Berkovich diamond indenter. Measurements were performed at room temperature. Prior to nanomechanical measurements, the system was calibrated, and the area function of the indenter tip was measured on fused silica. Nanohardness (H) and reduced Young's modulus (Er) values were extracted from the unloading data, applying the Oliver and Pharr model [27]. For each irradiation state, at least 15 measurements were made.

The first step of the nanoindentation study was to choose the best experimental conditions in order to probe the irradiated region without the impact of the unirradiated region (or at least with its minimal influence). One of the key test parameters is the maximum force, which defines penetration depth [28]. Optimum forces were selected so that the corresponding nanoindentation depths are as shallow as possible to minimize the effect of the unimplanted region being sampled and simultaneously probe material as deep as possible to limit effects such as not fully developed plastic zone [29], surface preparation [30] and ISE (Indentation Size Effect) [31]. As mentioned above, according to SRIM calculations, the depths of ion-modified layers for 250 keV and 1.2 MeV irradiations were 70 and 270 nm, respectively. Consequently, a maximum nanoindentation force of 0.4 mN was selected for low-energy irradiated samples. This force corresponds to plastic depths of around 30 nm, which represents approx. 40 % of the thickness of the irradiated layer. The mechanical property information obtained under such experimental conditions represents the response of both irradiated layer and unmodified volume (minimal contribution is expected to occur). A maximum nanoindentation force of 1 mN was selected for high-energy irradiated samples. This

**Accepted Version**
Publication date: July 2023
Embargo: 24 Months
European Union, Horizon 2020, Grant Agreement number: 857470 — NOMATEN — H2020-WIDESPREAD-2018-2020
DOI: https://doi.org/10.1016/j.nimb.2023.03.027

force corresponds to a plastic depth of around 40 nm, representing approx. 15 % of the thickness of the irradiated layer. In this regard, the effect of unmodified bulk on the obtained results is considered negligible. One should keep in mind that according to the 10 % rule, no effect of the steel substrate on the measured mechanical properties is anticipated. All measurements were performed in load-control mode with loading, unloading, and dwell times of 5, 3, and 1 s, respectively.

3. Results and discussion

Our work aims to compare findings of nanomechanical measurements conducted on implanted to different depths of thin amorphous ceramic coatings. Before proceeding to the nanoindentation measurements, the surface quality of the samples was evaluated. SEM images and average surface roughness values $R_a$ of PLD-grown alumina coating before and after ion irradiations are depicted in Fig. 3. One can observe that, regardless of the irradiation energy, no significant effect of implantation on the surface morphology can be observed. The coating is smooth, compact, and free of discontinuities.

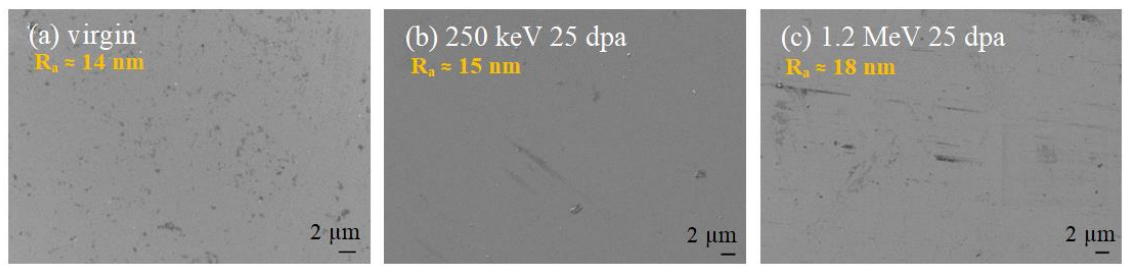

Fig. 3. SEM images of the surface of PLD-grown alumina coating: (a) virgin; (b) after 250 keV $Au^+$ irradiation up to 25 dpa; (c) after 1.2 MeV $Au^+$ irradiation up to 25 dpa.

Hardness and Young's modulus of alumina coating as a function of irradiation dose are shown in Fig. 4 and 5, respectively. Presented graphs compare the nanoindentation data obtained for low (250 keV) and high energy (1.2 MeV) irradiated samples. One can observe that the qualitative relationship between the nanohardness and damage level created by ion irradiations is comparable for both experimental energies. In both cases, measured hardness decreases with the damage level at the early stage of irradiation. At the 3 dpa, it shows a minimum and eventually slightly increases (or

Publication date: July 2023
Embargo: 24 Months
European Union, Horizon 2020, Grant Agreement number: 857470 — NOMATEN — H2020-WIDESPREAD-2018-2020
DOI: https://doi.org/10.1016/j.nimb.2023.03.027

becomes stable) for high dpa levels. Quantitatively speaking, the dose dependence of the nanohardness varies depending on the experimental design. It can be clearly seen that the hardness values measured for the low energy experiment are approximately 15 % lower. Attempts to explain the phenomenon in terms of nanoindentation issues are given below. In Fig. 5, load-displacement data for virgin and 3 dpa samples are selectively compared. The similarity between the graphs is noteworthy.

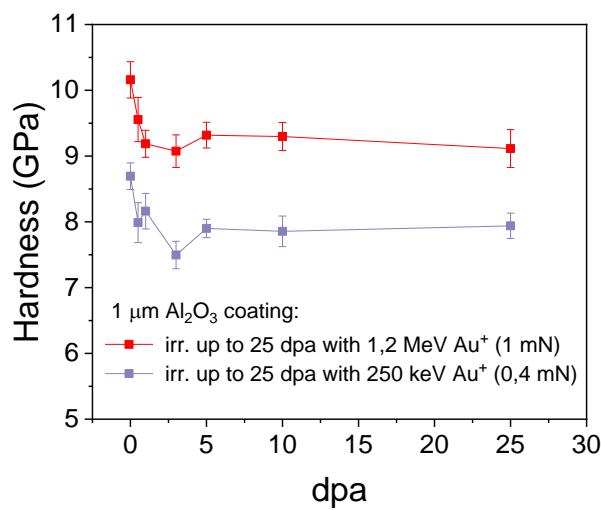

Fig. 4. Nanohardness in the function of dpa measured for PLD-grown $Al_2O_3$ coating irradiated with 250 keV (0.4 mN testing force) and 1.2 MeV (1 mN testing force). Both specimens were irradiated with Au ions. Adapted from [32] and [48].

**Accepted Version**
Publication date: July 2023
Embargo: 24 Months
European Union, Horizon 2020, Grant Agreement number: 857470 — NOMATEN — H2020-WIDESPREAD-2018-2020
DOI: https://doi.org/10.1016/j.nimb.2023.03.027

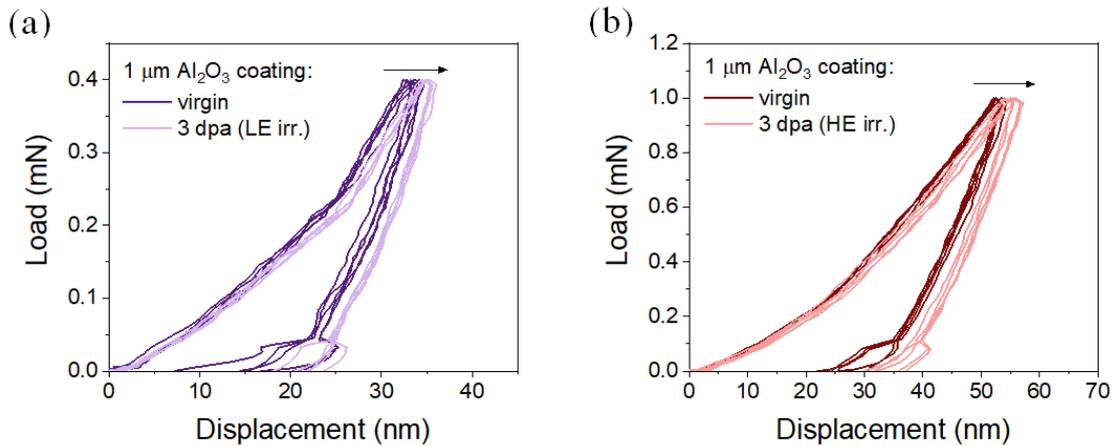

Fig. 5. Load-displacement behavior of PLD-grown $Al_2O_3$ coating before and after $Au^+$ irradiation (a) with 250 keV (0.4 mN testing force) (b) 1.2 MeV (1 mN testing force)

    To better understand the nanoindentation results for the irradiated samples, it is first necessary to look at the virgin data. As reported by F. García Ferré et al. [14], in the as-deposited state, amorphous PLD-grown alumina is characterized by the hardness of 10.3 ± 1.0 GPa. In this study, the hardness values of 8.7 ± 1.0 GPa and 10.2 ± 0.3 GPa were registered for the virgin sample tested at 0.4 and 1 mN forces, respectively. Hardness calculated from 1 mN force measurements is in complete agreement with the one reported by F. García Ferré et al. [30]. In contrast, when the material was tested with 0.4 mN force, extracted hardness values were found to be slightly undervalued. Registered underestimation is considered to be the result of the tip-rounding effect, as described by A. C. Fischer-Cripps [29]. A. C. Fischer-Cripps suggests that, for low penetration depths, the mean contact pressure (which is the measure of hardness) does not reflect the actual hardness of the film. It results from the fact that the indenter tip pressed into the material is initially round, so most of the deformation may be elastic. Generally, increasing indentation depth is necessary to overcome the issue. In the present case, however, this would result in losing information from the irradiated layer (for the low-energy irradiated sample). Thus, the testing force couldn't be increased. Consequently, all registered hardness values are underestimated for the low energy irradiated sample that was tested with 0.4 mN. We believe that the phenomenon described above is the source of the discrepancy when high and low energy data are compared in quality. This means that the

**Accepted Version**

Publication date: July 2023
Embargo: 24 Months
European Union, Horizon 2020, Grant Agreement number: 857470 — NOMATEN — H2020-WIDESPREAD-2018-2020
DOI: https://doi.org/10.1016/j.nimb.2023.03.027

observed quantitative mismatch does not result directly from different irradiation conditions but indirectly from using various nanoindentation testing forces. Moreover, it should be pointed out that the most pronounced tip-rounding behavior overlaps with the range of other undesired effects that can contribute to the actual data obtained. These include: (i) inhomogeneous dose profile, (ii) indentation size effect, (iii) implantation and surface effects. The quantitative analysis of such results should not be carried out. Accordingly, the critical issue is whether this kind of data may be interpreted qualitatively. Although the data on low energy irradiated material was affected by several effects in this research, the qualitative relationship between nanohardness and irradiation damage level obtained in low energy experiment is very similar to the correlation found in high energy experiment. This fact may suggest that indeed, such low-energy results may find valuable. Indeed, it should be regarded as a qualitative result. But it still may be helpful when the study aims to initially assess a material response to ion irradiation. For example, as a screening test procedure, as the first step before high energy experiment at high temperature.

As stated in the *Introduction*, ion irradiation brings many befits. However, despite many associated advantages, using an ion beam is not free from challenges. The first difficulty arises from the inhomogeneous dose profile. Damage level changes across the penetration depth, and the Bragg peak, characterized by the highest damage level, occurs very often. This can easily be seen on the damage profiles for irradiations performed in this work (Fig. 1). The second drawback of the method is that the volume of modified material is restricted to the near-surface region. In fact, the ion penetration depth depends on the type and energy of the ions but typically does not exceed a few microns [32,33]. Penetration of low-energy ions in the material can be very shallow (ten to hundreds of nanometers). This is also the case in this study, where for 250 keV ions irradiation depth is below 100 nm. The possibilities of extracting mechanical data (hardness or young modulus) from such a thin layer are extremely limited. In such cases, nanoindentation experiments may be the only route to obtain information on the mechanical behavior of the irradiated system. This method makes it possible to characterize small volumes.

**Accepted Version**

Publication date: July 2023
Embargo: 24 Months
European Union, Horizon 2020, Grant Agreement number: 857470 — NOMATEN — H2020-WIDESPREAD-2018-2020
DOI: https://doi.org/10.1016/j.nimb.2023.03.027

However, several challenges are associated with nanoindentation and the application of this technique to characterize ion irradiated materials [34]. The issues to be addressed include mainly: inhomogeneous dose profile, indentation size effect (ISE), unirradiated substrate effect, implantation, and surface effects [30,34,35]. These effects make mechanical estimation difficult, as was discussed for 250 keV experiment. For this reason, it is necessary to move detection limits further and provide as accurate data as possible, which helps to grasp and understand occurring phenomena. It is worth to point out, that the results of some studies suggest, that the commonly accepted limitation of indentation depth not larger than 1/10 of the layer thickness is excessive, and reliable data on layer nanohardness can be obtained even in case when the penetration depth reaches half of the layer thickness [36].

Considerations described above focus on comparing nanoindentation studies conducted for two energies (250 keV and 1.2 MeV). Regarding the overall influence of room temperature ion irradiation on the performance of PLD-grown alumina, this was the subject of another work [25]. In the paper, attempts were made to answer why the coating hardness reduction is observed after ion irradiation. Comprehensive structural characterization (XRD, TEM) conducted as a part of the research revealed that the as-deposited coating is amorphous and room temperature irradiation has no effect on the material structure. Based on the analysis of the mechanical and structural data obtained, it is considered that the most likely cause for the softening phenomenon lies with gold interstitials injected into the material during irradiation. This is because the indentation plastic zone extends deeper than the sample stopping-peak region, and the volume with chemically modified composition is probed. In general, reported room temperature radiation endurance results show no notable change in coating performance following irradiation to 25 dpa. All these findings demonstrate excellent properties of PLD-grown alumina coatings and encourage further development and qualification process.

4. Conclusions

Any material devoted to nuclear applications (like aforementioned coatings) must be appropriately qualified, and its behavior in operating conditions has to be fully

**Accepted Version**
Publication date: July 2023
Embargo: 24 Months
European Union, Horizon 2020, Grant Agreement number: 857470 — NOMATEN — H2020-WIDESPREAD-2018-2020
DOI: https://doi.org/10.1016/j.nimb.2023.03.027

understood. The combination of ion irradiation and nanoindentation has been recognized as an effective tool for predicting the material's response to neutron damage. In this study, we compared the nanomechanical information from low (250 keV) and high energy (1.2 MeV) Au+ irradiated thin alumina coatings. The obtained results show that the data obtained in the low-energy experiment may be considered reliable, pointing to the conclusion that in some cases, the commonly known 10% rule may be obeyed. Interestingly, although these data were affected by several effects (usually considered detrimental for this kind of study), the qualitative relationship between nanohardness and irradiation damage level obtained in the low and high-energy experiments is very similar. This observation suggests that qualitative assessment of the mechanical changes using nanoindentation may be feasible even for very shallow implantation depths. Since shallow irradiation results in uniform development of the radiation damage zone (no commonly visible Bragg peak) obtained result demonstrates that in some cases, one can build a credible research methodology by using low-energy ion implanter programs (which generally are more available than high-energy ion labs). The discussed approach gives a qualitative estimate of the expected changes. This may be considered beneficial when access to a high-energy ion source is poor, and pre-estimation of the material's radiation tolerance must be done. However, one should remember that extreme caution must be taken when interpreting such results. Hence this methodology can be applied only for very hard materials with very low surface roughness.

Acknowledgments

The research leading to these results was carried out in the frame of EERA Joint Programme on Nuclear Materials and is partly funded by the European Commission Horizon 2020 Framework Programme under grant agreement No. 755269 (GEMMA project). We acknowledge support from the European Union Horizon 2020 research and innovation program under grant agreement no. 857470 and from the European Regional Development Fund via the Foundation for Polish Science International Research Agenda PLUS program grant No. MAB PLUS/2018/8. Also, financial support from the National Centre for Research and Development through a research grant "Studies of the role of

**Accepted Version**

Publication date: July 2023
Embargo: 24 Months
European Union, Horizon 2020, Grant Agreement number: 857470 — NOMATEN — H2020-WIDESPREAD-2018-2020
DOI: https://doi.org/10.1016/j.nimb.2023.03.027
interfaces in multi-layered, coated and composite structures" PL-RPA2/01/INLAS/2019 is gratefully acknowledged. The Research Council of Norway is acknowledged for the support to the Norwegian Micro- and Nano-Fabrication Facility, NorFab, project number 295864. Lastly, financial support from the Ministry of Science and Higher Education granted under agreement no. 3906/H2020-EURATOM/2018/2 is gratefully acknowledged.

References

[1]   A. Alemberti, M. Caramello, M. Frignani, G. Grasso, F. Merli, G. Morresi, M. Tarantino, ALFRED reactor coolant system design, Nucl. Eng. Des. 370 (2020) 110884. doi:10.1016/j.nucengdes.2020.110884.

[2]   G. Grasso, C. Petrovich, D. Mattioli, C. Artioli, P. Sciora, D. Gugiu, G. Bandini, E. Bubelis, K. Mikityuk, The core design of ALFRED, a demonstrator for the European lead-cooled reactors, Nucl. Eng. Des. 278 (2014) 287–301. doi:10.1016/j.nucengdes.2014.07.032.

[3]   D. Iadicicco, M. Vanazzi, F.G. Ferré, B. Paladino, S. Bassini, M. Utili, F. Di Fonzo, Multifunctional nanoceramic coatings for future generation nuclear systems, Fusion Eng. Des. 146 (2019) 1628–1632. doi:10.1016/j.fusengdes.2019.03.004.

[4]   G. Balakrishnan, P. Kuppusami, S.T. Sundari, R. Thirumurugesan, V. Ganesan, E. Mohandas, D. Sastikumar, Structural and optical properties of γ-alumina thin films prepared by pulsed laser deposition, Thin Solid Films. 518 (2010) 3898–3902. doi:10.1016/j.tsf.2009.12.001.

[5]   E. Miorin, F. Montagner, V. Zin, D. Giuranno, E. Ricci, M. Pedroni, V. Spampinato, E. Vassallo, S.M. Deambrosis, Al rich PVD protective coatings: A promising approach to prevent T91 steel corrosion in stagnant liquid lead, Surf. Coatings Technol. 377 (2019) 124890. doi:10.1016/j.surfcoat.2019.124890.

[6]   J. Gottmann, E.W. Kreutz, Pulsed laser deposition of alumina and zirconia thin films on polymers and glass as optical and protective coatings, Surf. Coatings
© <2023>. This manuscript version is made available under the CC-BY-NC-ND 4.0 license
https://creativecommons.org/licenses/by-nc-nd/4.0/

**Accepted Version**

Publication date: July 2023
Embargo: 24 Months
European Union, Horizon 2020, Grant Agreement number: 857470 — NOMATEN — H2020-WIDESPREAD-2018-2020
DOI: https://doi.org/10.1016/j.nimb.2023.03.027

**Accepted Version**

Publication date: July 2023
Embargo: 24 Months
European Union, Horizon 2020, Grant Agreement number: 857470 — NOMATEN — H2020-WIDESPREAD-2018-2020
DOI: https://doi.org/10.1016/j.nimb.2023.03.027

**Accepted Version**

Publication date: July 2023
Embargo: 24 Months
European Union, Horizon 2020, Grant Agreement number: 857470 — NOMATEN — H2020-WIDESPREAD-2018-2020
DOI: https://doi.org/10.1016/j.nimb.2023.03.027

**Accepted Version**

Publication date: July 2023
Embargo: 24 Months
European Union, Horizon 2020, Grant Agreement number: 857470 — NOMATEN — H2020-WIDESPREAD-2018-2020
DOI: https://doi.org/10.1016/j.nimb.2023.03.027

**Accepted Version**

Publication date: July 2023
Embargo: 24 Months
European Union, Horizon 2020, Grant Agreement number: 857470 — NOMATEN — H2020-WIDESPREAD-2018-2020
DOI: https://doi.org/10.1016/j.nimb.2023.03.027